\documentclass[letterpaper,twocolumn,aps]{revtex4-1}

\usepackage{graphicx}
\usepackage{color}
\definecolor{Red}{rgb}{1,0,0}

\begin{document}

\title{On the Reconstructed Fermi Surface in the Underdoped Cuprates}

\author{H.-B. Yang,$^{1}$ J. D. Rameau,$^{1}$ Z.-H. Pan,$^{1}$ G. D. Gu,$^1$ P. D. Johnson,$^{1}$\\
 H. Claus,$^2$ D. G. Hinks,$^2$ and T. E. Kidd$^{3}$} \affiliation{
(1) Condensed Matter Physics and Materials Science Department, Brookhaven National Laboratory, Upton, New York 11973, USA \\
(2) Materials Science Division, Argonne National Laboratory, Argonne, Illinois 60439, USA\\
(3) Physics Department, University of Northern Iowa, Cedar Falls, Iowa 50614, USA }

\begin{abstract}
The Fermi surface topologies of underdoped samples of the high-$T_c$ superconductor
Bi2212 have been measured with angle resolved photoemission. By examining thermally
excited states above the Fermi level, we show that the observed Fermi surfaces in the
pseudogap phase are actually components of fully enclosed hole pockets. The spectral
weight of these pockets is vanishingly small at the anti-ferromagnetic zone boundary,
which creates the illusion of Fermi ``arcs'' in standard photoemission measurements. The
area of the pockets as measured in this study is consistent with the doping level, and
hence carrier density, of the samples measured. Furthermore, the shape and area of the
pockets is well reproduced by phenomenological models of the pseudogap phase as a spin
liquid.
\end{abstract}
\maketitle

Understanding the pseudogap regime in the high $T_c$ superconducting cuprates is thought
to be the key to understanding the high $T_c$ phenomenon in general \cite{Timusk}.   An
important component of that understanding will be the determination of the nature of the
low lying normal state electronic excitations that evolve into the superconducting state.
It is therefore critically important to know the exact nature of the Fermi surface (FS)
associated with these materials.  Photoemission studies of the pseudogap regime reveal
gaps in the spectral function in directions corresponding to the copper-oxygen bonds and
a FS that seemingly consists of disconnected arcs falling on the surface defined within
the framework of a weakly interacting Fermi liquid \cite{MRNorman}. A number of different
theories have attempted to explain these phenomena in terms of competing orders whereby
the full FS undergoes a reconstruction reflecting the competition
\cite{Chubukov,Chakravarty}.  An alternative approach recognizes that the superconducting
cuprates evolve with doping from a Mott insulating state with no low energy charge
excitations to a new state exhibiting properties characteristic of both insulators and
strongly correlated metals.

Several theories have been proposed to describe the cuprates from the latter perspective
\cite{XGWen98,YRZ06,QYang10}. One such approach is represented by the so-called YRZ
ansatz \cite{YRZ06}, which, based on the doped RVB spin liquid concept
\cite{PAnderson87}, has been shown to successfully explain a range of experimental
observations in the underdoped regime \cite{KYYang09,Leblanc10A,Borne10A,Leblanc10B}. The
model is characterized by two phenomena, a pseudogap that differs in origin from the
superconducting gap and hole-pockets that satisfy the Luttinger sum rule for a Fermi
surface defined by both the poles and zeros of the Green's function at the chemical
potential \cite{KRT06}. The pockets manifest themselves along part of the FS as an
``arc'' possessing finite spectral weight corresponding to the poles of the Green's
function as in a conventional metal. The remaining``ghost'' component of the Fermi
surface is defined by the zeros of the Green's function and therefore posses no spectral
weight to be directly observed. Importantly, the zeros of the Green's function at the
chemical potential coincide with the magnetic zone boundary associated with the
underlying antiferromagnetic order of the Mott insulating state and therefore restrict
the pockets to lying on only one side of this line. The model further predicts that the
arc and ghost portions of the FS are smoothly connected to pockets. Several theoretical
studies indicate that within this framework the pockets have an area that scales with the
doping \cite{YRZ06,Leblanc10A}.  Recent photoemission studies have indeed provided some
indication that the pseudogap regime is characterized by hole pockets centered in the
nodal direction \cite{HBYangNat08,JMeng09}.  In the present study, we demonstrate for the
first time that the FS of the underdoped cuprates in the normal state is characterized by
hole pockets with an area proportional to the doping level and a Fermiology as described
above. These measurements, well fit by the YRZ Green's function, show that the FS and
associated properties of high-$T_c$ cuprate superconductors need not rely on the
existence of exotic phenomena such as disconnected Fermi``arcs'' for quantitative
explanation.

The photoemission studies reported in this paper were carried out on underdoped cuprate
samples, both Ca doped and oxygen deficient.  The Ca-rich crystal was grown from a rod
with Bi$_{2.1}$Sr$_{1.4}$Ca$_{1.5}$Cu$_2$O$_{8+\delta}$  composition using an arc-image
furnace with a flowing 20\% O$_2$/Ar gas mixture.  The maximum $T_c$ was 80 K. The sample
was then annealed at 700 $^{\circ}$C giving a 45 K $T_c$ with a transition width of 2 K.
The oxygen deficient Bi$_2$Sr$_2$CaCu$_2$O$_{8+\delta}$ crystals were produced by
annealing optimally-doped Bi-2212 crystals,  at 450 $^{\circ}$C to 650 $^{\circ}$C for 3
$\sim$ 15 days. The spectra shown in this paper were all recorded on beamline U13UB at
the NSLS using a Scienta SES2002 electron spectrometer. Each spectrum was typically
recorded for a period of five to six hours in the pulse-counting mode with an energy and
angular resolution of 15 meV and 0.1$^{\circ}$ respectively.

\begin{figure}[htbp]
\begin{center}
\includegraphics[width=8cm]{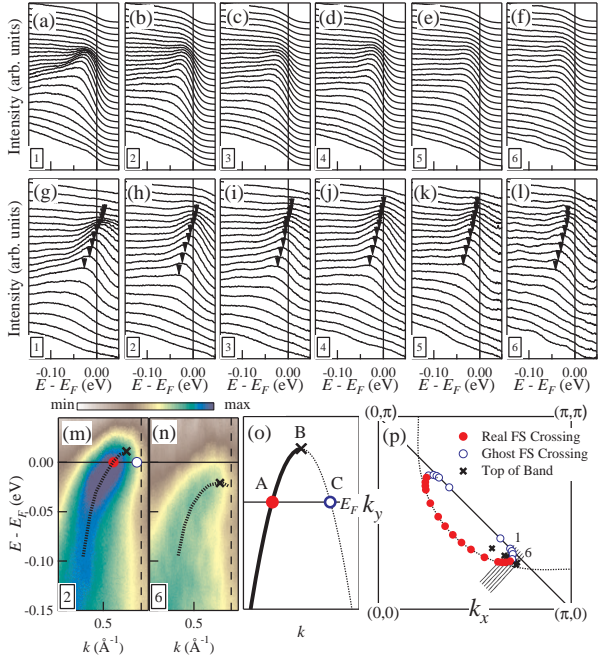}
\caption{Fermi surface of under-doped Bi2212 at 140 K ($T_c$ = 65 K). (a-f) EDCs of raw
data taken at cuts 1-6 in (p), and (g-h) after the analysis described in the text. The
EDC peak positions are marked with triangles. In cuts 1-6, the band is slowly moving down
with respect to $E_F$. Fermi surface crossing points exist in cuts 1-4, but not in cuts 5
\& 6. These cuts demonstrate how the band sinks below $E_F$ at the edge of the Fermi
``pocket''. (m) Image plot of cut 2 after Fermi normalization, with the FS crossing (red
circle) and the top of the band (black cross) indicated. The ``ghost'' FS crossing (blue
open circle) deduced from symmetrization in momentum is also indicated. (n) The same
image plot for cut 6 showing no FS crossing. (o) Schematic showing points analyzed in the
measured spectra as indicated in the text. (p) Pseudo-pocket determined for the 65 K
sample.  Red circles indicate the measured FS crossings corresponding to point A in the
schematic. Crosses show the measured extremity of the dispersion corresponding to point
B, and the open circles represent the ``ghost'' FS corresponding to point C. The dashed
line indicates the large LDA FS. } \label{1}
\end{center}
\end{figure}

Figure \ref{1} (a-f) shows photoemission spectra obtained near the end of the measured
``Fermi arc'' for an underdoped Bi2212 ($T_c$ = 65 K) sample.  The spectra in Fig.
\ref{1} (g-l) are shown after analysis using the Lucy-Richardson (LR) deconvolution
approach \cite{JRameauLucy} to reduce the effects of the experimental resolution and
after division by the appropriate temperature dependent Fermi function.  This approach
allows a more accurate determination of the FS crossings.  The spectrum in Fig.
\ref{1}(m) contains two experimental observations as indicated in the schematic in Fig.
\ref{1}(o), the directly measured FS crossing, point A, and the point at which the
dispersion comes to an abrupt halt, point B.  We associate the point B with a gap in the
spectral function reflecting the scattering of the photohole in the underlying spin
liquid. As noted earlier, several calculations indicate that the formation of a Fermi
hole pocket reflecting a particle-hole asymmetry in binding energy is derived from this
scattering \cite{XGWen98,YRZ06}. Alternative models that recognize the strong
correlations in the system can also produce pockets \cite{QYang10, Granath10}. Within the
YRZ ansatz, the pocket is formed by two band crossings, points A and C in the schematic,
symmetric about the zone point associated with the bottom of the gap derived from the
scattering, point B. Thus if we observe one FS crossing, point A, and the bottom of the
gap, point B, we can in principle determine the other side of the pocket, the ``ghost''
FS, point C, as indicated in Fig. \ref{1}(p). The Fermi pocket derived in this manner is
compared in the figure with the full FS traditionally assumed in ARPES studies.  The
deviation between latter FS and that determined in the present study becomes most evident
near the end of the``arc''.  It is important to note that the hole pocket determined in
this manner is clearly asymmetric with respect to the magnetic zone boundary, ruling out
any pockets generated by scattering mechanisms simply associated with a $Q(\pi,\pi)$
vector. The ``ghost'' portion of the FS is, however, consistent with models showing a
surface of zeros in the Green's function at the chemical potential running along the
magnetic zone boundary, defined by the line from $(0,\pi)$ to $(\pi,0)$
\cite{YRZ06,Konik06}. Under the condition, $G(\vec{k},\omega)=0$, the spectral weight
measurable by photoemission is vanishingly small  because the spectral function is
defined by $A(\vec{k},\omega)= - \textrm{Im} G(\vec{k},\omega )$.

\begin{figure}[htbp]
\begin{center}
\includegraphics[width=8cm]{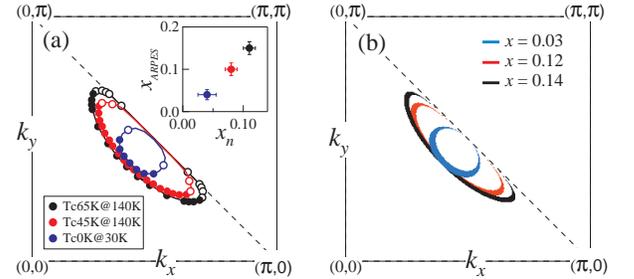}
\caption{(a) The pseudo-pockets determined for three different doping levels.  The black
data corresponds to the $T_c$ = 65 K sample, the blue data corresponds to the $T_c$ = 45
K sample and the red data corresponds to the non-superconducting $T_c$ = 0 K sample. The
area of the ``pockets'' $x_{ARPES}$ scales with the nominal of doping level $x_n$, as
shown in the inset. (b) The Fermi pockets derived from YRZ ansatz with different doping
level.} \label{2}
\end{center}
\end{figure}

Having determined the approximate shape and size of the pocket we can calculate the
associated hole density for a given sample. Assuming the area inside the magnetic zone
boundary corresponds to one electron at half filling, the pocket area corresponds to a
hole carrier density of 0.15, higher than the doping level determined from the measured
$T_c$ alone. However the measured area is in reasonably good agreement with the area
bounded by the locus of superconducting Bogoliubov band minima $k_B(E)$ extracted from
SI-STS studies of a BSSCO sample with a similar doping level \cite{Kohsaka08}.   As such
the combination of the two experiments raises obvious questions.  Does the pocket area
really scale with the doping level and how is that consistent with earlier studies
suggesting the ``arc'' length is temperature dependent with a length proportional to
$T/T^*$ \cite{Kanigel}  (Here $T^*$ represents the doping dependent pseudogap temperature
scale.) In attempting to answer these questions we show in Figure \ref{2}(a)  a
comparison of the FSs obtained using the present approach for the 65 K sample, a Ca doped
sample ($T_c$ = 45 K) and an oxygen-deficient non-superconducting sample ($T_c$ = 0 K).
It is clear from Fig. \ref{2}(a) that reducing the doping level into the highly
underdoped regime results in a more noticeable deviation from the ``LDA'' FS. Further,
while the measured areas of the different pockets, 0.15 holes ($T_c$ = 65 K) 0.13 holes
($T_c$ = 45 K) and 0.04 holes ($T_c$ = 0 K) are larger than the presumed doping levels,
0.11, 0.085 and $<$ 0.05 respectively. It is clear that the pocket size scales in
relationship to the doping level as predicted theoretically \cite{YRZ06,Leblanc10A}.  The
finding of a finite nodal FS rather than a ``nodal'' point at low $T$ for the $T_c$ = 0 K
sample is at variance with recently reported findings under the same conditions
\cite{Chatterjee10}.  The measured Fermi pockets are however in good agreement with those
predicted by the YRZ ansatz. In Fig. \ref{2}(b) we show the spectral function calculated
at $E_F$ as a function of doping, where $A(\vec{k},0) = -(1/\pi
)\textrm{Im}G^{YRZ}(\vec{k},0)$ and where
\begin{equation}
G^{YRZ}(\vec{k},\omega) =
{g_t\over{\omega-\xi_0(\vec{k})-\Delta^2_R/{[\omega-\xi_0(\vec{k})]}}},
\end{equation}
with associated parameters taken from reference \cite{YRZ06}. The shape and area of the
FS measured in ARPES are remarkably well reproduced by this model with the doping level
as the only adjustable parameter.

\begin{figure}[htbp]
\begin{center}
\includegraphics[width=7cm]{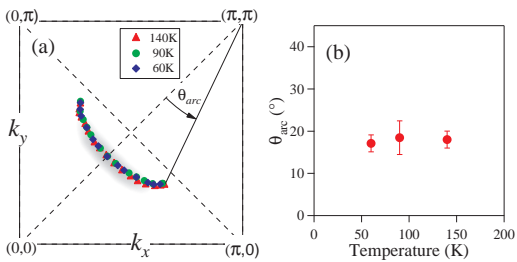}
\caption{(a) The Fermi surface crossings determined for the $T_c$ = 45 K sample at three
different temperatures.  The triangles indicate measurements at a sample temperature of
140 K, the circles measurements at 90 K and the diamonds measurements at 60 K.  (b) the
measured arc lengths in (a) plotted as a function of temperature. We note that rather
than cycling the temperatures on the same sample, the data in (a) are measured on
different samples cut from the same crystal. } \label{3}
\end{center}
\end{figure}

Turning to the question of whether the pocket areas are temperature dependent we show in
Figure \ref{3}(a) the observed Fermi ``arc'' for the $T_c$ = 45 K sample measured at
three different temperatures: 60 K, 90 K and 140 K,  all in the normal state but well
below $T^*$. The measured FS crossings in the figure are determined by the same method
used in Figs. \ref{1} and \ref{2} rather than from the spectral weight at the Fermi
level. In Fig. \ref{3}(b) we show the measured arc length as a function of temperature.
It is clear that any change with temperature is minimal and certainly not consistent with
an increase by more than a factor of two between the data taken at 140 K and 60 K as
would be expected by a $T/T^*$ scaling of the ``arc'' length \cite{Kanigel}. This
discrepancy arises because several experiments showing that the length of the Fermi
``arc'' scales as $T/T^*$ depend upon an examination of the spectral weight at the
chemical potential rather than the direct determination of whether or not a band actually
crosses the chemical potential.

The picture of the low energy excitations of the normal state emerging from the present
study is of a nodal FS characterized by a Fermi ``pocket'' that, at temperatures above
$T_c$, shows a minimal temperature dependence and an area proportional only to the doping
level. The bounds of this pocket are formed by smoothly connected surfaces of poles and
zeros of the Green's function well described by the YRZ ansatz. To fill out this picture
of the normal state we turn our attention to the antinodal pseudogap itself.

\begin{figure}[htbp]
\begin{center}
\includegraphics[width=8cm]{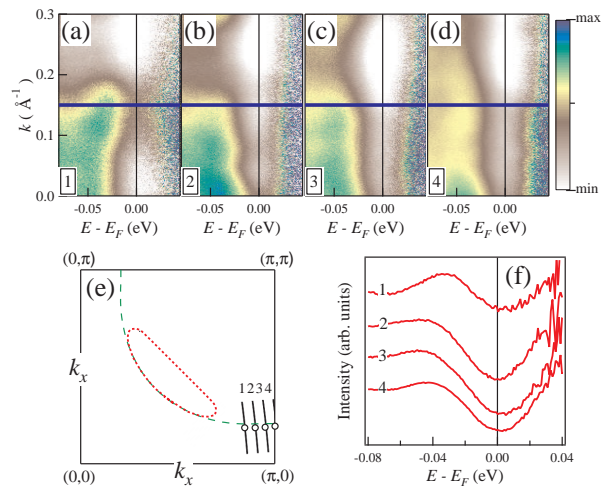}
\caption{(a-d) Spectra intensity measured at four points in the anti-nodal region as
indicated in the schematic (e).  The measurements are made with the sample in the normal
state at a temperature of 140 K.   (f) Intensity cuts through the spectral intensity maps
of panel (a-d) as indicated by the lines in panel (a-d).  They are also indicated by the
open circles in (e).} \label{4}
\end{center}
\end{figure}

Several theories of the pseudogap phase propose the formation of preformed singlet pairs
above $T_c$ in the anti-nodal region of the Brillouin zone \cite{EmeryKivelson95}.  The
YRZ spin liquid based on the RVB picture is one such model as it recognizes the formation
of resonating pairs of spin singlets along the copper-oxygen bonds of the square lattice
as the lowest energy configuration. Fig. \ref{4}(a-d) shows a series of spectral plots
along the straight sector of the ``LDA'' FS in the anti-nodal region at a temperature of
140 K for the $T_c$ = 65 K sample at the locations indicated in Fig. \ref{4}(e).  Fig.
\ref{4}(f) shows intensity cuts through these plots along the horizontal lines indicated
in Fig. \ref{4}(a-d). It is evident that a symmetric gap exists at all points along this
line. The particle-hole symmetry in binding energy observed here is in marked contrast to
the particle-hole symmetry breaking predicted in the presence of density wave order and
is a necessary condition for the formation of Cooper pairs.   Thus the present
observations add support to the hypothesis that the normal state is characterized by pair
states forming along the copper-oxygen bonds and is consistent with earlier studies.

The combination of Fig. \ref{2} and Fig. \ref{4} points to a more complete picture of the
low energy excitations in the normal state of the underdoped cuprates.  For $T_c < T <
T^*$, a Fermi ``pocket'' exists in the nodal region with an area proportional to the
doping level. One does not need to invoke discontinuous Fermi ``arc''s to describe the FS
of underdoped Bi2212 and Luttinger's sum rule, properly understood, is seen to still
approximately stand. Although not verified in the present study one assumes that at some
critical doping level greater than optimal, the reconstructed FS switches to the full
Fermi surface as predicted in calculations \cite{YRZ06,Leblanc10A}. The full FS ought to
also be visible for $T> T^*$, as has been observed recently in ARPES measurements
\cite{Hashimoto10}. In the underdoped regime, as one moves away from the pockets, two
further distinct regions exist as manifested in the spectral intensity. The region in the
immediate vicinity of the end of the pockets is characterized by a gap that is asymmetric
with respect to the chemical potential. This gap again reflects the underlying spin
correlation in the system.  In the antinodal region, the gap becomes symmetric with
respect to the chemical potential and is therefore indicative of incoherent preformed
pairs of electrons in singlet states. This picture of the FS appears to be entirely
consistent with the momentum dependence of the gap function found in STS studies of the
same material \cite{Pushp09}.

In conclusion, the photoemission measurements presented here show that the FS of the
underdoped Bi2212 cuprate superconductors is defined not by disconnected ``arcs'', but
instead by fully closed hole pockets with areas proportional to the doping level and thus
the carrier density of the system. The size and shape of these pockets are well
reproduced within phenomenological models of the pseudogap state, such as the YRZ spin
liquid, and are consistent with the doping in a Mott insulator. The absence of spectral
weight on the ``ghost'' side of the pockets along the magnetic zone boundary results from
the fact that quasiparticles in this region are defined by zeros, not poles, of the
single particle Green's function. These results show the cuprates evolve with doping from
an antiferromagntetic insulator into a pseudogap state characterized by a heavily
renormalized band structure and strong pairing correlations reflecting the underlying
structure of the lattice. These are the essential characteristics of the underdoped
cuprates, the unraveling of which is prerequisite to understanding the ultimate emergence
of superconductivity below $T_c$.

The authors would like to thank Seamus Davis, Mike Norman, Maurice Rice, John Tranquada,
Alexei Tsvelik, Subir Sachdev, Tonica Valla and Ali Yazdani for useful discussions.  The
work at Brookhaven is supported in part by the US DOE under Contract No.
DE-AC02-98CH10886 and in part by the Center for Emergent Superconductivity, an Energy
Frontier Research Center funded by the US DOE, Office of Basic Energy Sciences.  The work
at Argonne is partially supported by the Department of Energy under contract no.
DE-AC02-06CH11357 and partially supported by the same Center for Emergent
Superconductivity contract. T. E. Kidd acknowledges support from the Iowa Office of
Energy Independence grant No. 09-IPF-11.

\end{document}